# Atomic-like behaviors and orbital-related Tomonaga-Luttinger liquid in peapod quantum dots

J.Mizubayashi[1,5], J.Haruyama[1,5], I.Takesue[1,5], T.Okazaki[2,5], H.Shinohara[3,5], Y.Harada[4,5], Y.Awano[4,5]

[1] Aoyama Gakuin University, 5-10-1 Fuchinobe, Sagamihara, Kanagawa 229-8558 Japan
[2] National Institute of Advanced Industrial Science and Technology, Tsukuba, 305-8565, Japan
[3] Nagoya University, Furo-cho, Chigusa, Nagoya  464-8602  Japan
[4] Fujitsu Laboratory, 10-1 Wakamiya, Morinosato, Atsugi, Kanagawa  243-0197 Japan,
[5] JST-CREST, 4-1-8 Hon-machi, Kawaguchi, Saitama 332-0012 Japan

**We report ballistic charge transport phenomena observed in carbon-nanotube peapod quantum dots. We find atomic-like behaviors (shell filling) sensitive to applied back-gate voltages ($V_{bg}$) by single electron spectroscopy. Doubly degenerate electronic levels are found only around ground states at very low $V_{bg}$. Those correlations with presence of nearly free electron states unique to the peapods are discussed. Moreover, we find power laws in conductance versus energy relationships with anomalously high values of power, which are strongly associated with shell filling to the doubly degenerate levels. It is investigated that the powers originate from Tomonaga-Luttinger liquid via the occupied doubly degenerate levels. These results imply that a ballistic charge transport is still preserved at low $V_{bg}$ regions in peapod quantum dots in spite of presence of the encapsulated $C_{60}$ molecules.**

Nano-peapods, which are single-walled carbon nanotubes (SWNTs) encapsulating a series of fullerenes, such as $C_{60}$, $C_{70}$, and $Gd@C_{82}$ ($C_{82}$ encapsulating Gd) in their inner space [1, 2], have recently attracted considerable attention. This is because their remarkable nanostructures yield exotic electronic states, charge transports, and one-dimensional (1D) quantum phenomena. However, there are still a few reliable reports that reported those electronic states and quantum phenomena.

From theoretical viewpoints, in $C_{60}@(n,n)$ peapods that are arm-chair type SWNTs encapsulating $C_{60}$ molecules, it has been predicted that electrons that were transferred from the SWNT accumulated in the space between the $C_{60}$ molecules and SWNTs, forming the so-called nearly free electron (NFE) states [3]. Hybridization of these NFE states with the π and σ orbitals of $C_{60}$ introduced four asymmetric subbands including the approximately doubly degenerate ground states in the $C_{60}@(10,10)$ peapod in contradiction to the two subbands in conventional SWNTs [3, 4].

Measurements of semiconductive peapods encapsulating a series of $Gd@C_{82}$ by a scanning tunnel microscope revealed that a conduction band was periodically modulated around $Gd@C_{82}$ in a real space due to the hybridization of orbitals between the SWNT and $Gd@C_{82}$ [5]. Moreover, electrical measurements of peapods encapsulating $C_{60}$ and $Gd@C_{82}$ indicated the possibility of the presence of variable range hopping [6]. Refs. [3 – 6] at least suggested the presence of charge transfer and orbital hybridization between the encapsulated fullerenes and SWNTs.

On the other hand, it is well known that SWNTs are within a 1D ballistic charge transport regime and exhibit a variety of quantum effects, such as quantized energy levels, Tomonaga-Luttinger liquid (TLL) [7 – 10], and atomic-like behaviors (shell filling) as quantum dots [11 – 14]. For instance, the behavior of TLL, which is a collective phenomenon arising from electron-electron interaction in 1D conductors, has been identified by observing power laws in relationships of conductance vs. energy in carbon nanotubes (CNs) [7 – 10]. The reported correlation exponent $g$, which denotes the strength of an electron-electron interaction, was as low as ~ 0.2. This implied the presence of a strong repulsive Coulomb interaction in CNs. When CNs act as quantum dots, electron can be placed on the quantized electronic levels in the dots one by one due to single charging effect. This effect has caused atomic-like behaviors in CN quantum dots [7 - 14], such as even-odd effect, shell-filling in two spin-degenerate electronic states, and Kondo effect. How such phenomena are affected by encapsulating a series of fullerenes, however, has not yet been investigated in any carbon nano-peapods to date.

For the present study, field-effect transistors (FETs) using peapods encapsulating $C_{60}$ molecules as the channel were fabricated. An SEM top view indicated that the FETs included two bundles of peapods [2] as the channels. The number of peapods included in one bundle was estimated to be approximately 20 from measurements by the SEM, AFM, and TEM. Since the observed differential conductance was largely independent of the change in back gate voltage ($V_{bg}$), metallic transport in the present peapod was suggested [20].

First, the measurement results by single electron spectroscopy are shown in Fig.1(a) and Fig.2. Figure 1(a) shows Coulomb diamonds observed in the $V_{bg}$ region < +2V. The four diamonds, a sequence of one large diamond (shown as n = 4) followed by three smaller ones (shown as n = 1 – 3), are observable. This sequence indicates possible presence of atomic-like behaviors with the doubly degenerate electronic levels only at ground states, based on previous reports of the four fold diamonds in SWNT [13] and multi-walled CN (MWNT) quantum dots [14]. However, the observed sequence of these diamonds was only one set only at $0V < V_{bg} < +2V$. This result is very different from those periodically observed over wide ranges of $V_{bg}$ in refs.[13, 14]

In conventional CN quantum dots, such one-set degenerate levels cannot exist, because individual non-degenerate electronic level is formed only from quantization of two subbands existing in bulk of a SWNT, while only in some cases all levels are doubly degenerate like the four fold diamonds as observed in refs. [13] and [14].

Hence, in order to confirm presence of doubly degenerate levels for Fig.1(a), we have investigated the $V_{bg}$ shift of the linear-response conductance peaks (i.e., shown by arrows in Fig.1(a)) as a function of magnetic field, B, perpendicular to the tube axis. The result, Fig.1(b), reveals that adjacent peaks shift in opposite directions. This is a behavior of spin singlet state whose spins alternate as S=0 →1/2 →0 ……. and exist on the same orbital state, unlike a spin triplet state formed by Hund's rule. In Fig.1(c), we plot addition energy, which was deduced from the separation of adjacent peaks involving electrons on the same orbital in Fig.1(b) (i.e., peaks 1 and 2, peaks 3 and 4), as a function of magnetic field, B. A dashed line shows the result of the best fit of the data to $U_c + g_L\mu_B B$, where $\mu_B$ is the Bohr magneton and $g_L$ is the Lande factor, and gives $g_L = 1.96$. This value of $g_L$ is approximately consistent with those mentioned in ref.[14]. Therefore, we conclude that Fig.1(a) indicates presence of doubly degenerate electronic levels existing only at ground states and presence of atomic-like behaviors (shell filling).

We interpret that these one-set doubly degenerate levels originate from the encapsulated $C_{60}$ molecules and NFE states unique to peapods from the following six reasons;

**1.** Our empty SWNTs quantum dots, in which $C_{60}$ molecules are not encapsulated with the same structures as those used for the present peapods, exhibited just even-odd effects (i.e., non-degenerate discrete levels) as shown in Fig.2(a) with $\Delta E = \sim 3$ meV, which is consistent with the previous studies in SWNT quantum dots (e.g., a $\Delta E$ of ~5 meV for a tube length of 100 ~ 200 nm) and the relationship $\Delta E = hv_F/2L$ (L and $v_F$ are the tube length and Fermi velocity, respectively). This obviously stresses that the observed doubly degenerate levels are unique to peapod quantum dots in our case in contradiction to refs.[13,14] and, hence, the encapsulated $C_{60}$ molecules yield the degenerate levels.

**2.** The observed atomic-like behaviors were very sensitive to applied $V_{bg}$ as follows. The doubly degenerated levels disappeared and only even-odd effect appeared in the $V_{bg}$ region higher than that in Fig.1(a) (i.e., at $+2V < V_{bg} < +5V$) as shown in Fig.2(b). Moreover, at $V_{bg} > +5$ V, this even-odd effect disappeared and only conventional Coulomb diamonds appeared. Furthermore, in entire -$V_{bg}$ region, no atomic-like behavior has been detected and only Coulomb diamonds appeared. Because we have confirmed this tendency in three samples at least, these are not accidental results like stochastic Coulomb diamonds reported in some previous papers.

**3**. The value of $U_c$ (the single-electron charging energy of the system) is approximately 6 ~ 10 meV in the small diamonds in Fig.1(a). This value for the present peapod with a length of 500 nm is approximately consistent with expectations based on a previous study of SWNT bundles (e.g. a $U_c$ of ~25 meV for a tube length of 100–200 nm) [11]. Hence, this result stresses that the encapsulated $C_{60}$ molecules do not contribute to the effective electrostatic capacitance ($C_e$) for the single charging effect of the peapod quantum dot in this low $V_{bg}$ region < +2V.

**4.** In contrast, the sizes of Coulomb diamonds in Fig.2(b) become smaller than those in Fig.1(a). This means a decrease in $U_c$ to ~2 meV from 6 ~ 10 meV due to an increase in $C_e$ and that the encapsulated $C_{60}$ molecules are electrostatically coupled with the SWNT in parallel for $C_e$ in this $V_{bg}$ region; $+2V < V_{bg} < +5V$. The value of capacitance of $C_{60}$ molecules are estimated to be three ~ five times larger than that in the SWNT.

**5.** In contradiction to these results, it is a well known fast that conventional CN quantum dots can place many electrons on those many quantized electronic levels one by one via single electron tunneling and show periodical atomic-like features over wide $V_{bg}$ regions, because the shape of dots are independent of $V_{bg}$ and $V_{bg}$ just changes positions of chemical potentials in the CN quantum dots unlike most of semiconductor quantum dots.

6. Ref.[3] predicted that the doubly degenerate subbands existing only at the ground states originated from the hybridization of orbitals in $C_{60}$ molecules and NFE states in bulk of $C_{60}$@(10,10) peapod as mentioned in introduction. In the case of quantum dot structure, however, these subbands should result in quantized electronic levels. In this sense, presence of the one-set doubly degenerate electronic levels observed only around ground states in Fig.1(a) is qualitatively relevant, because of non-modulation of $C_{60}$ molecules by very low $V_{bg}$ (< +2V).

Based on the 1st term, the 2nd – 5th terms stress that the applied high $V_{bg} > 2V$ modulates the encapsulated $C_{60}$ molecules and change the shapes of Coulomb diamonds in the present peapod quantum dots. If the encapsulated $C_{60}$ molecules have no chemical bonds to the SWNT and can freely rotate inside the SWNT, this modulation cannot occur. Hence, this means presence of chemical bonds between $C_{60}$ molecules and SWNT and, therefore, also presence of NFEs even in the present peapod quantum dot as ref [3] predicted in bulk of peapods. This implies that $V_{bg}$ modulates not only $C_{60}$ molecules but also NFEs at Vbg > +2V. Consequently, the doubly degenerate levels observed in the $V_{bg} < +2V$ can be attributed to non-modulated NFEs. This is qualitatively consistent with ref.[3] as explained in the 6th term.

When applied $V_{bg}$ increases to the region in $+2V < V_{bg} < +5V$, electrons start to be accumulated on the SWNT. Because the NFEs had been formed by electron transfer from the SWNT, this accumulation indicates that NFEs are transferred back to the SWNT, resulting in those depletions. This depletion of NFEs modulates electronic levels, leading to elimination of one-set doubly degenerate levels. However, non-degenerate electronic levels and atomic-like behaviors still survive under coupling with $C_{60}$ molecules for single electron charging effect, resulting in Fig.2(b). Further increase in $V_{bg}$ to the region $> +5V$ makes the NFEs entirely deplete and hybrid states disappear. This leads to disappearance of atomic-like behaviors. Because applying $-V_{bg}$ causes electron depletion in the SWNT and electron transfer from the SWNT to NFEs, absence of any atomic-like behaviors in the entire $-V_{bg}$ region indicates that such an electron transfer are sensitive much more than those in $+V_{bg}$ region.

Consequently, it can be confirmed that peapod quantum dots can preserve a ballistic charge transport in spite of presence of the encapsulated $C_{60}$ molecules and NFEs, when a small $+V_{bg}$ (e.g., $V_{bg} < +5V$ in the present case) was applied. Moreover, it should be noticed that the applied low $+V_{bg}$ (e.g., $V_{bg} < +2V$ in the present case) play a major role for oreserving NFEs without modulation and observation of atomic-like behaviors as similar as those in bulk peapods.

Next, behaviors of orbital-related TLL states are discussed. Figure 3 shows the double-logarithmic plot of differential conductance divided by $T^{\alpha}$ as a function of $eV/k_BT$ measured at $V_{bg} = +0.4V$ for three different temperatures. All data collapse on a single universal value with showing a saturation at $eV/k_BT < hv_F/L$. This result stresses presence of TLL states in the present peapod [8], but showing a larger value of $\alpha$. Figure 4 shows the relationships of differential conductance ($dI_{sd}/dV_{sd}$) to source-drain voltage ($V_{sd}$) on doubly logarithmic scales for one - $V_{bg}$ and three + $V_{bg}$ regions. In the $-V_{bg}$ region, any differential conductance did not follow a linear relationship as shown in Fig.4(a). On the contrary, saliently linear relationships with different $\alpha$ values are observable in the $+V_{bg}$ region. The behaviors are classified into three regions (Fig.4 (b) - (d)) as mentioned in the figure captions, showing anomalously large values of $\alpha$ ($1.6 < \alpha < 12$).

The summary of values of $\alpha$ observed in all the $V_{bg}$ region included in Fig.4(b) - (d) are shown in Fig.4(e). The differences in tendencies of $\alpha$ among the three regions are apparent in this figure. Moreover, the values of $\alpha$ observed in empty SWNTs used for Fig.1(b) are also shown in this figure. All the values are less than 1, which is consistent with previous reports of TLLs in SWNTs. This implies that Fig.4(a) is unique to peapods.

The presence of power laws has been discussed as evidence for TLLs in CNs [7 – 10], as mentioned in the introduction. The values of $\alpha$ were very sensitive to the boundary conditions between the metal electrodes and CNs [8], namely the tunneling density of state; such as $\alpha^{bulk}$ = ~ 0.3 and $\alpha^{end}$ = 2 $\alpha^{bulk}$ for the tunneling from an Au electrode to the bulk and to the end of CNs within the large-channel number TLL states, respectively [8]. The formulas of α for each tunneling were also given by $\alpha^{bulk}$ = ($g^{-1}$ + g – 2)/8 and $\alpha^{end}$ = ($g^{-1}$ – 1)/4. However, it should be noted that even the maximum value of α reported in CNs to date is approximately 1.25, except for refs.[9, 16]. Therefore, we imply that the α values of 1.6 ~ 12 observed in Figs.3 and 4 are anomalously large in comparison with the α values reported thus far in conventional TLLs [9]. The junction structures in this study, in which the ends of the peapod bundles were placed under an Au electrode, should have shown a maximum $\alpha^{end}$ of only ~0.6. In fact, the empty SWNTs have exhibited α = ~0.8 even at the maximum case as explained for Fig.4(e) above.

Here, it should be noted that the power laws shown in Fig.4 (b) – (d) exist at each $V_{bg}$ in the gray areas, which are just the nearest outside regions of the Coulomb diamonds shown in Fig.1(a). The three $+V_{bg}$ regions shown in Fig. 4(b) - (d) and Fig.4(e) are in good agreement with the three $V_{bg}$ regions classified by the boundaries of diamonds and, hence, the number of electrons confined in the dot as shown in Fig.1(a). These results clearly indicate that the power law behaviors with the large values of α are strongly associated with the number of (partially) occupied electronic levels, *N*, in each diamond

The correlation of power laws and α (TLLs) with the electronic-state filling effect (i.e., orbital filling effect) in CNs has not yet been reported in previous studies. Only a single study [17], however, predicted that a small *g* and large α could be obtained from the large *N* in peapods. The theory predicted $g = (1+2Nv_q/\pi \hbar v_F)^{-1/2}$ for armchair CNs, where *N* and $v_q$ are the number of (partially) occupied symmetric subbands with degenerate Fermi vector waves and the same band width, and the electron-electron interaction matrix element, respectively. If the subbands are asymmetric and each of them crosses the Fermi level only once, *N* can be replaced by *N*/2. This holds true for the subbands of the $C_{60}$@(10,10) peapod in this study.

We quantitatively examine the validity of this theory for the present measurement by replacing *N* to the number of electronic states and using the same value of $v_q$. The value of g = 0.135 is obtained from $\alpha^{end}$ = ($g^{-1}$ – 1)/4 [8] using α = 1.6 that is observed in region I (N=2). The value of $v_q$ can be estimated by substituting these values of *g* and N in $g = (1 + 2(N/2)v_q/\pi \hbar v_F)^{-1/2}$[17]. Then, g = 0.11 and 0.099 are respectively obtained for N = 3 and N = 4 by substituting the estimated value of $v_q$ in $g = (1 + 2(N/2)v_q/\pi \hbar v_F)^{-1/2}$. The value of g = 0.11 for N = 3 is approximately in good agreement with g = 0.082 estimated from $\alpha^{end}$ = ($g^{-1}$ – 1)/4 by using α = 2.8 that is observed in the portion of region II with low $V_{sd}$ values.

On the other hand, this *g* value is irrelevant to α = 8 ~ 10 that is observed in the portion of region II with high $V_{sd}$ values. Moreover, the *g* value of 0.099 leads to α = 2.28 for N = 4, which is significantly less than the values of α > 10 observed in the region III at high $V_{sd}$ values. These indicate that different values of $v_q$ should be used for the case of higher $V_{sd}$. Because strength of electron-electron interaction varies from low to high $V_{sd}$s, this is reasonable. When different values of $v_q$ are used for large $N$, $\alpha = 10$ (for N = 3) and 12 (for N = 4) could be obtained from the values of $2v_q/\pi\hbar v_F$ = 1160 and 1250, respectively, $g = (1 + 2(N/2)v_q/\pi\hbar v_F)^{-1/2}$, and $\alpha^{end} = (g^{-1} - 1)/4$.

Consequently, the theory [17] is quantitatively relevant when N = 2 and 3 (at lower $V_{sd}$) under the same value of $v_q$ and N = 3 (at higher $V_{sd}$) and 4 under the larger values of $V_q$. This indicates that the presence of two power laws observed in Fig.4(c) is attributed to change in $v_q$ due to increase in $V_{sd}$. Therefore, we conclude that the power laws with large values of $\alpha$ ($1.6 < \alpha < 12$) can be attributed to the TLL via the occupied doubly degenerated electronic levels, which are located near the ground states unique to the peapod quantum dots. Here, presence of TLL states means also presence of a ballistic charge transport. Because power laws were not detected in $-V_{bg}$ region, this is consistent with absence of the atomic-like behaviors in $-V_{bg}$ region. Further investigation is, however, required to reconfirm the values of $v_q$. As shown in Fig.4(c)(d), the $V_{sd}$ regions exhibiting power laws are very narrow at $V_{sd}$'s > 10 mV, i.e., at best half decade and the high values of power do not become constant in the region III. Hence, other interpretation may be possible like [21].

The observations reported in this study strongly suggest that the 1D quantum phenomena observable in peapods are very exotic and sensitive to the encapsulated $C_{60}$ molecules, NFE states, applied $V_{bg}$, and shell filling via the NFE states. Further investigation is required in order to develop a comprehensive understanding of these phenomena (e.g., changing the number of encapsulated $C_{60}$ molecules).

We acknowledge T.Nakanishi, S.Tarucha, W.Izumida, and M.Thorwart for valuable discussions

environment shown for MWNTs [8] may explain the $\alpha = 8 \sim 12$, because $\alpha$ in conventional MWNTs is given by $2R/R_Q = 2(L/C)^{1/2}/R_Q \approx 0.44$, where L is the kinetic inductance given by $R_Q/2Nv_F$ ($\approx 1nH/\mu m$), C is the external electrostatic capacitance ($\approx 30$ aF/$\mu$m), and $R_Q$ ($=h/e^2$) is the quantum resistance.

**Figure captions**

**Fig.1: (a)** Coulomb diamonds (white regions surrounded by the dotted lines) observed in a peapod quantum dot at $V_{bg}$ < +2V and at T = 1.5 K. The z-axis is the differential conductance with the magnitudes of which are indicated on the right side. *n* indicates the number of electrons confined in each diamond. $V_{bg}$ regions I, II, and III well correspond to those in Fig.4(e).
**(b)** $V_{bg}$ shift of conductance peak positions (at $V_{bg}$= 0.11 V, 0.26 V, 0.53 V, and 0.77 V around $V_{sd}$ = 0V shown by arrows in (a)) in Fig.1(a) as a function of magnetic field, *B*.
**(c)** Addition energy obtained from each peak pair in (b) versus *B*.

**Fig.2: (a)** Coulomb diamonds observed around ground states in an empty SWNT quantum dot (i.e., without encapsulating $C_{60}$ molecules) with the same structures as that used for Fig.1.
**(b)** Typical Coulomb diamonds measured at +5V > $V_{bg}$ > +2V and at T = 1.5 K in the peapod used for Fig.1

**Fig.3:** The double-logarithmic plot of differential conductance ($G = dI_{sd}/dV_{sd}$) divided by $T^\alpha$ as a function of $eV/k_BT$ measured at $V_{bg}$ = +0.4V for three different temperatures.

**Fig.4:** Relationships of $dI_{sd}/dV_{sd}$ to source-drain voltage ($eV_{sd}$ /k >> T = 1.5 K measured) on doubly logarithmic scales for four different $V_{bg}$ regions. These power laws primarily appear just the nearest outside regions of diamonds (i.e., in the gray areas as shown by arrow) of Fig.1(a). Only the power law in the low $V_{sd}$ region at $V_{bg}$ = 1 V appears in the n = 4 diamond. The liner lines were obtained from accurate data fitting including measurement points as many as possible and values of power α were exactly estimated. **(a):** For $-V_{bg}$ region. No dI/dV follows a linear relationship. **(b) - (d):** For three $+V_{bg}$ regions. **(b):** The linearities with 1.6 < α < 2 are observable only at $V_{sd}$ < 0.01 V. **(c):** Two linear relationships with different α ranges (i.e., α = 2 ~ 3 and α = 8 ~ 10 for $V_{sd}$ < 0.01 V and $V_{sd}$ > 0.02 V, respectively) are observable. **(d):** The linearities with α = 10 ~ 12 are observable only at $V_{sd}$ > 0.01 V.
**(e)** Dependence of power α on different $V_{bg}$ values, estimated from Fig.4(b)-(d), in present peapod and empty SWNT quantum dots. Three $V_{bg}$ regions (**(I):** $V_{bg}$< 0.8 V, **(II):** 0.8 V < $V_{bg}$ < 1.8 V, and **(III):** 1.8 V < $V_{bg}$ corresponding to Fig.4 (b), (c), and (d), respectively) are evident. Several values of α were added in addition to Fig.4(d) only in region III.

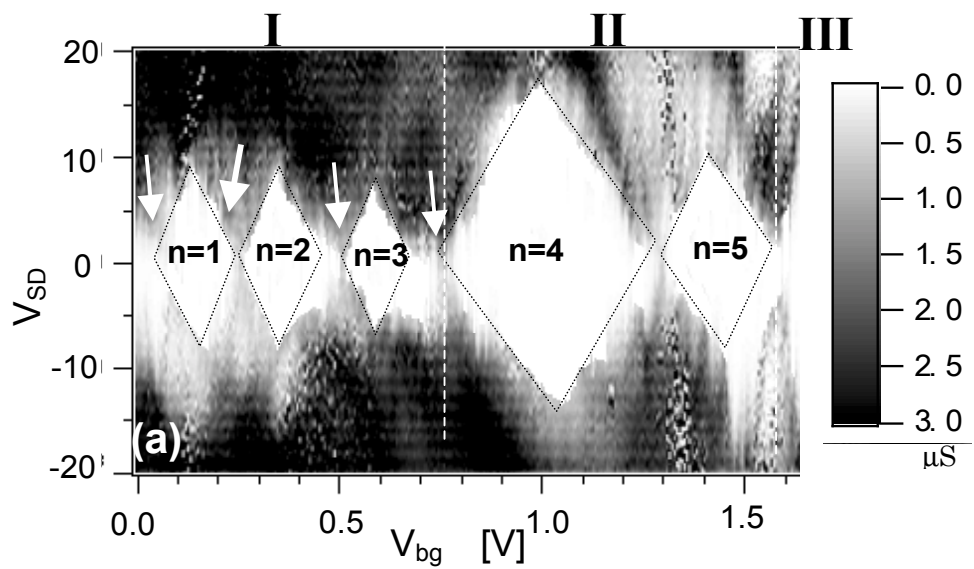

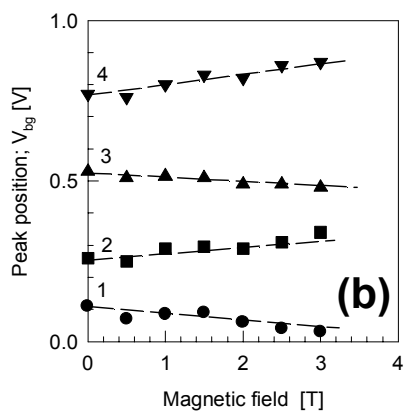
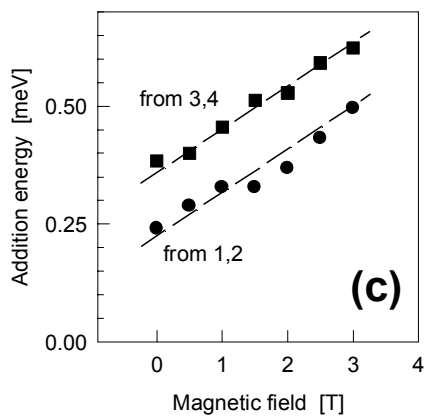

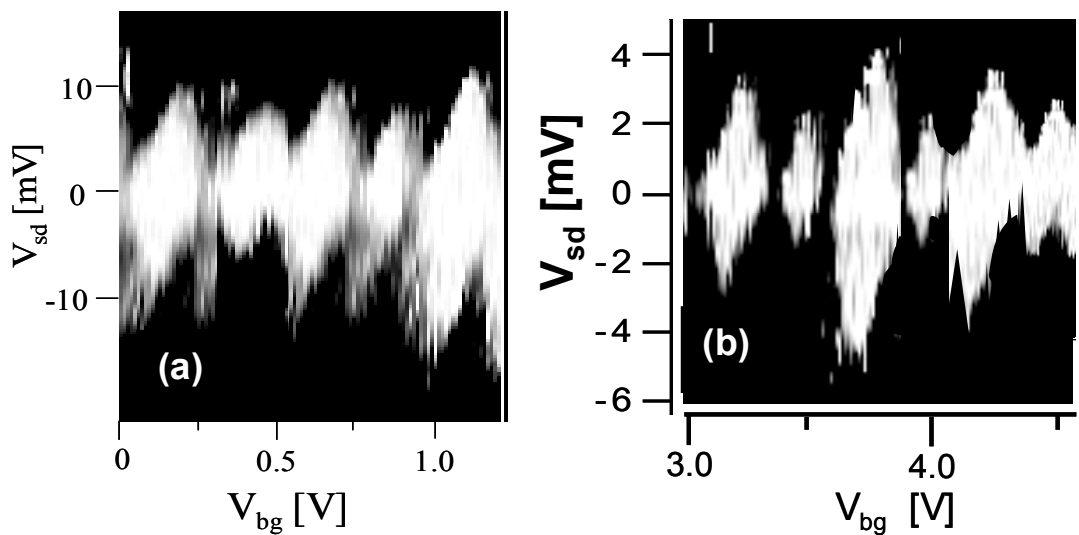

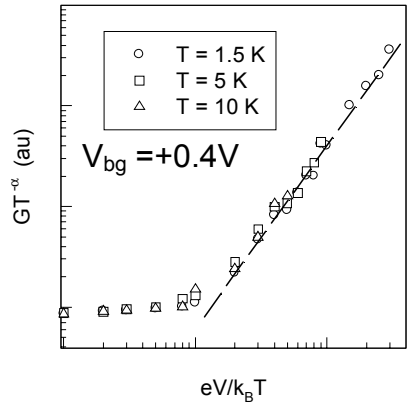

**Fig.3**

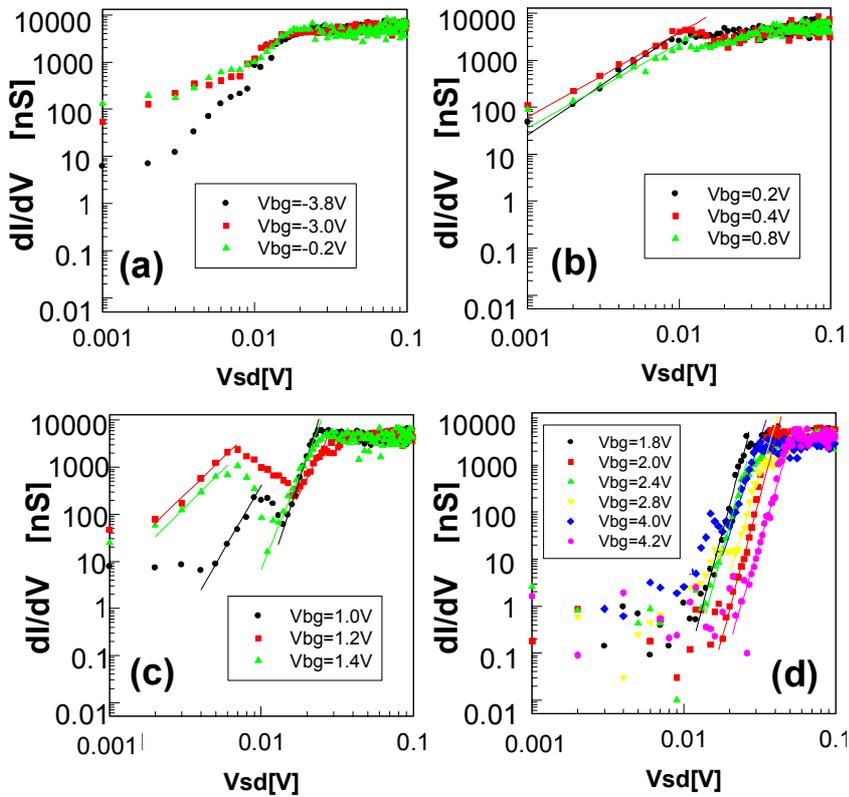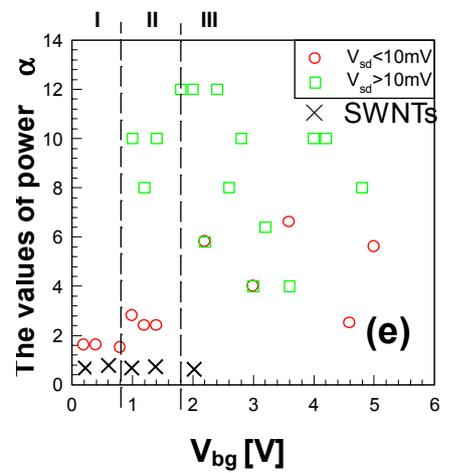

**Fig.4**